\documentclass[11pt]{iopart}
\usepackage{graphicx}
\usepackage{subfigure}
\usepackage{amssymb}
\usepackage{epstopdf}
\usepackage{color}
\usepackage[square, comma, numbers, sort&compress]{natbib}

\begin{document}

\title{Un-modeled search for black hole binary systems in the NINJA project}
\author{Laura Cadonati$^1$, Shourov Chatterji$^{2,3}$, Sebastian Fischetti$^1$,
Gianluca Guidi$^{2,4}$, Satyanarayan~R.~P.~Mohapatra$^1$, Riccardo Sturani$^{2,4}$, Andrea Vicer\'e$^{2,4}$}
\address{$^1$  Department of Physics, University of Massachusetts, Amherst, MA 01003}
\address{$^2$ INFN-Sezione Firenze, I-50019 Sesto Fiorentino, Italy}
\address{$^3$ LIGO -- California Institute of Technology, Pasadena, CA 91125, USA}
\address{$^4$ Istituto di Fisica, Universit\`a di Urbino, I-61029 Urbino, Italy}

\ead{\mailto{cadonati@physics.umass.edu}}

\begin{abstract}
The gravitational wave signature from binary black hole coalescences is an important target for ground based interferometric detectors such as LIGO and Virgo.  
The Numerical INJection Analysis (NINJA) project brought together the numerical relativity and gravitational wave data analysis communities, with the goal to optimize the detectability of these events.
In its first instantiation, the NINJA project produced a simulated data set with numerical waveforms from binary black hole coalescences of various morphologies (spin, mass ratio, initial conditions), superimposed to Gaussian colored noise at the design sensitivity for initial LIGO and Virgo.
We analyzed the NINJA simulated data set with the {\em Q-pipeline} algorithm, designed for the all-sky detection of gravitational wave bursts with  minimal assumptions on the shape of the waveform. 
The algorithm filters the data with a bank of sine-Gaussians, sinusoids with Gaussian envelope, to identify significant excess power in the time-freuqency domain. 
We compared the performance of this burst search algorithm with {\em lalapps\_ring}, which match-filters data with a bank of ring-down templates to specifically target the final stage of a coalescence of black holes. 
A comparison of the output of the two algorithms on NINJA data in a single detector analysis yielded qualitatively consistent results; however due to the low simulation statistics in the first NINJA project, it is premature to draw quantitative conclusions at this stage, and further studies with higher statistics and real detector noise will be needed.
\end{abstract}


\titlepage

\section{The NINJA Project}

The Numerical INJection Analysis (NINJA) project~\cite{Aylott:2009ya,Cadonati:2009} is a collaboration of 
numerical relativists and gravitational-wave data analysts, with the goal to improve detectability of binary black hole (BBH) coalescences with numerical relativity waveforms. The first NINJA project represented an important step towards the use of numerical waveforms to enhance the performance of data analysis.
Nine numerical relativity groups shared waveforms for BBH coalescences, with no restrictions on spin, eccentricity, or mass ratio, and ten data analysis groups analyzed them with various methods based on both modeled (i.e. matched-filtering) and un-modeled searches.
Numerical relativity can now provide complete coalescence waveforms, from \emph{inspiral} to \emph{ring-down} of the final remnant, through the \emph{merging} of the black hole constituents of the binary system. 
These waveforms offer a unique opportunity to test the data analysis search pipelines for efficiency in detection and faithfulness in source parameter estimations.
A recent review of the status of binary black hole simulations is available in~\cite{Hannam:2009rd}.

The NINJA project added numerical relativity BBH waveforms to Gaussian noise {\em colored} by multiplying the flat spectrum by the power spectral density of each of the Laser Interferometer Gravitational-wave Observatory (LIGO)~\cite{Abbott:2007kv,Smith:2009} and  Virgo~\cite{Acernese2006} detectors, at design sensitivity. 
This ideal, simulated  Gaussian noise did not include typical features of real data, such as non-Gaussian noise transients and narrow band features like the mirrors' oscillation normal modes or {\em violin modes}, the suspension thermal noise around the frequency of standing wave modes of the  mirror suspensions.
A population of simulated gravitational signals has also been produced, from numerical relativity data. Such population covers a broad range of black hole masses, distances and orientations.
For the numerical waveforms that were used, 
see~\cite{Brugmann:2008zz,Husa:2007hp,% BAM
Alcubierre:2000xu,Alcubierre:2002kk,Koppitz:2007ev,Pollney:2007ss,% CCATIE
Imbiriba:2004tp,vanMeter:2006vi,% Hahndol
Zlochower:2005bj,Campanelli:2005dd,% LazEv
Sperhake:2006cy,% Lean
Hinder:2007qu,% MayaKranc
Pretorius:2004jg,Pretorius:2005gq,% PU
Scheel:2006gg,% SpEC
Etienne:2007hr% UIUC
}, for descriptions of the numerical codes 
see~\cite{Hannam:2007ik,Hannam:2007wf,% BAM_HHB
Tichy:2008du,% BAM_FAU
Pollney:2007ss,Rezzolla:2007xa,% CCATIE 
% Hahndol
% LazEv
% Lean
Vaishnav:2007nm,Hinder:2007qu,% MayaKranc
Buonanno:2006ui,Pretorius:2007jn,% PU
Boyle:2007ft,Scheel:2008rj,% SpEC
Etienne:2007hr% UIUC
}.
For additional details on this data set, and a description of the parameter space covered by these waveforms, we refer to~\cite{Aylott:2009ya,Cadonati:2009}.

The NINJA project analyzed this data set with several analysis techniques, developed and used by the LIGO Scientific Collaboration (LSC) and Virgo data analysis: (1) matched filtering with analytical waveform models of the inspiral stage,  from Post-Newtonian expansion~\cite{CDFnrda}, (2) matched filtering with hybrid models of the full coalescence, created by matching Post-Newtonian perturbation templates with numerical relativity waveforms~\cite{AEInrda}, (3) matched filtering to ring-down templates, to target the final phase of the coalescence, with the {\em lalapps\_ring} code (Sec.~\ref{s:ring}) and (4) un-modeled searches that do not rely on templates, with two algorithms, the {\em Q-pipeline} (Sec.~\ref{s:qpipe}) and Hilbert Huang Transform~\cite{HHTnrda}. 
Parameter estimation techniques were also tested on this data set~\cite{BayesNRDA,MCMCnrda}. Details and a comparisons of the performance of all methods in NINJA is provided in~\cite{Aylott:2009ya}.

This paper aims to highlight the potential of un-modeled analysis in the detection of BBH coalescences, by comparing the performance of the 
 {\em Q-pipeline}~\cite{ChatterjiThesis,Chatterji2004,omega} algorithm to {\em lalapps\_ring}~\cite{lalapps,Creighton} on the first NINJA simulated data set.  
Both algorithms are applied to the output of a single LIGO simulated detector, whose data span a duration of $\sim$30 hours. Sec.~\ref{analysis} describes the analysis, Sec.~\ref{results} presents the results, and conclusions are drawn in Sec.~\ref{conclusions}.

%~~~~~~~~~~~~~~~~~~~~~~~~~~~~~~~~~~~~~~~~~~~~~~~~~
\section{The Analysis}\label{analysis}
%~~~~~~~~~~~~~~~~~~~~~~~~~~~~~~~~~~~~~~~~~~~~~~~~~
Vacuum solutions to the Einstein's equation for high mass systems are scale invariant, and the total mass of the system determines the frequency scale for the waveform: the higher is the mass, the smaller is the frequency of the coalescence, and the fewer cycles are in the sensitive band of ground-based interferometers like LIGO and Virgo.
For BBH coalescences in the mass range covered by the NINJA simulations (50--350~$\mbox{M}_{\odot}$), most of the signal-to-noise ratio is in the merger and ringdown phases~\cite{PhysRevD.57.4535}. 
Matched filter to ring-down templates and un-modeled searches sensitive to the merger are useful techniques to detect these systems; here we provide some detail on these searches, with reference to two specific algorithms used by the LSC and Virgo.

%~~~~~~~~~~~~~~~~~~~~~~~~~~~~~~~~~~~~~~~~~~~~~~~~~
\subsection{Matched filtering to ring-downs with {\em lalapps\_ring}}
\label{s:ring}
%~~~~~~~~~~~~~~~~~~~~~~~~~~~~~~~~~~~~~~~~~~~~~~~~~

The {\it lalapps\_ring}~\cite{lalapps,Creighton} code was developed by the LSC for ring-down searches. 
The code implements the matched filtering technique~\cite{Wainstein:1962} with a bank of damped sinusoids, to represent the quasi-normal modes of a black hole relaxation after the merger. 
For this study of NINJA data, we used the same version as in the LSC S4 ring-down analysis~\cite{GogginThesis,Abbott:2009km}. The template bank is made of
sinusoids with given frequency damped with an exponential, covering the parameter range 50-2000 Hz in frequency and 2-20 in Q$_\mathrm{rd}$, where Q$_\mathrm{rd}$ is $\pi$ times the damping time, in unit of the sinusoid period.
The analysis is performed in overlapping blocks of 2176 seconds; the low-frequency cutoff is 45 Hz for the analysis of LIGO data. The Virgo noise curve allows a lower cutoff, however in this paper we only report single detector results from the LIGO noise curve.

%~~~~~~~~~~~~~~~~~~~~~~~~~~~~~~~~~~~~~~~~~~~~~~~~~
\subsection{Un-modeled algorithms: {\em Q-pipeline}}\label{s:qpipe}
%~~~~~~~~~~~~~~~~~~~~~~~~~~~~~~~~~~~~~~~~~~~~~~~~~
A {\em gravitational wave burst} is a short duration signal for which the waveform is not necessarily known. The LSC and Virgo have implemented searches for such signals that do not use templates but instead look for instances of significant excess power in multiple detectors. Several algorithms have been developed, each making use of a different method for the creation of time-frequency maps and different details in the statistical analysis~\cite{chatterji-2004-21, TFclusters,Klimenko04,cWB,CorrPower,BN,HHTGW,XP,Searle}; in particular, the  {\em Q-pipeline}~\cite{ChatterjiThesis,Chatterji2004} is used  in the LSC-Virgo flagship search for gravitational wave bursts~\cite{burstS5}.
The {\em Q-pipeline} is a multi-resolution time-frequency search for statistically significant excess signal energy, equivalent to a templated matched filter search for sine-Gaussians in whitened data.
The template bank is constructed to cover a finite region in the following parameters:
(1) central time, or the time of maximum of the Gaussian envelope, 
(2) central frequency, that is the frequency of the sinusoid, and 
(3) quality factor Q$_\mathrm{q}$, or the number of oscillations under the gaussian envelope, which, up to a constant, is the ratio of central frequency to bandwidth of the signal. 
The mismatch between any sine-Gaussian in this signal space and the nearest basis function does not exceed a maximum of 20\% in energy.
For this study, as in ~\cite{burstS5}, the data is analyzed in 64 sec blocks, in the frequency range 48-2048 Hz, with  Q$_\mathrm{q}$ between 3.3 and 100.

%~~~~~~~~~~~~~~~~~~~~~~~~~~~~~~~~~~~~~~~~~~~~~~~~~
\subsection{Analysis strategy}
%~~~~~~~~~~~~~~~~~~~~~~~~~~~~~~~~~~~~~~~~~~~~~~~~~

Our ultimate goal is a multi-detector analysis, inclusive of a coherent followup that checks for sky location, and an event-by-event comparison of triggers from inspiral, burst and ring-down analyses, to explore the three phases of the coalescence, using {\em Q-pipeline}, {\em lalapps\_ring} and inspiral matched filtering triggers produced by one of the other NINJA analysis teams~\cite{Aylott:2009ya}.

\begin{figure}[htb]
  \centering
  \includegraphics[width=0.49\linewidth]{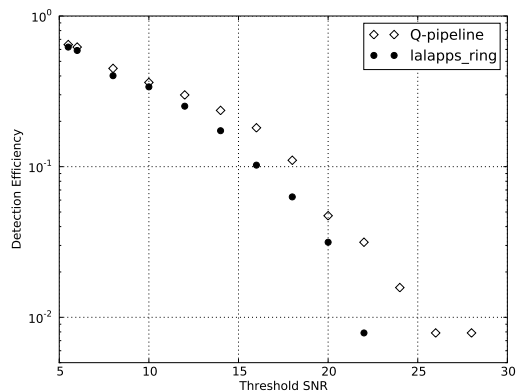}
  \caption{Single-detector efficiency for the population of signals in the NINJA data set for both algorithms, as a function of the SNR threshold. The left-most data point (SNR=5.5) is the threshold used in this analysis. }
  \label{fig:efficiency}
\end{figure}

However, for the first NINJA release, we forewent the coherent followup step and instead focused on single-detector results: we established a nominal signal-to-noise ratio (SNR) threshold of 5.5 and compared the parameters estimated by the two algorithms to source parameters: the time of the waveform maximum~$T_\mathrm{peak}$, and the innermost stable circular orbit (ISCO) and ring-down frequencies~$f_\mathrm{ISCO}$ and~$f_\mathrm{ring}$~\cite{Kidder,Echeverria,Goggin,Leaver}:
\renewcommand{\arraystretch}{1.6}
\begin{equation}
\label{e:freq}
\begin{array}{rcl}
f_\mathrm{ISCO} &=&\displaystyle \frac{c^3}{6\sqrt 6\pi G(m_1 + m_2)}\,,\\
f_\mathrm{ring} &=&\displaystyle \frac{c^3}{2\pi G M} \left[1-0.63(1-a)^{0.3}
\right]
\end{array}
\end{equation}
\renewcommand{\arraystretch}{1}
where~$G$ is the Newton constant, $c$ the speed of light, $m_{1,2}$ the individual constituent masses and $a,M$ are the final black hole dimension-less spin and mass calculated as  in~\cite{Rezzolla:2007rz,Rezzolla:2007rd} and~\cite{Buonanno:2007pf}, respectively. 
%These frequencies are compared to the frequency detected by both algorithms in Fig.~\ref{fig:freqQ}.
The injected SNR is computed from the signal before injection and the detector noise spectrum with starting frequency as specified in Sec.~\ref{s:ring}. 
Thresholds are consistent with the most recent published searches by the LSC-Virgo for ring-down~\cite{Abbott:2009km} and bursts~\cite{burstS5}. 
The dependence of the detection efficiency {\em for this set of simulations} is shown in Fig.~\ref{fig:efficiency}, for both algorithms; a more accurate threshold selection requires a study of accidentals, coincidence between multiple detectors and a fine tuning that goes beyond the scope of this initial study. 

\section{Results}
\label{results}
Unless otherwise stated, all the results discussed in the following sections are restricted to the 4 km Hanford detector (H1): given the limited statistics (94 signals injected with SNR$\ge 5.5$), a discussion of the response of all interferometers and a coincidence analysis would not reliably provide further insight in addition to the information presented below.

\subsection{Detection Performance}

The NINJA project paper~\cite{Aylott:2009ya} provides a comparative analysis of the 
different methods employed (inspiral, hybrid and ring-down  
templates for matched filtering, {\em Q-pipeline} and Hilbert Huang transform for
burst searches).
Here we focus on the analysis performed by the {\em Q-pipeline} un-modeled search 
 and by the {\em lalapps\_ring} matched-filtering to ring-down templates, both at 
the single interferometer level, with the same nominal threshold of 
SNR$_\mathrm{measured}\ge 5.5$.  
The statistics of this sample is too small to make inferences on which 
pipeline performs better in which parameter region; a more systematic study is 
needed to assess the power of the methods.  
Also, this analysis does not take into account the effect of background noise 
transients and accidental coincidences, which are very different in real data 
than in Gaussian noise, so this comparison is not complete.  Nevertheless we 
have an indication that all pipelines have comparable chances to find these 
signals.

\begin{figure}[htb]
  \centering
  \includegraphics[width=0.49\linewidth]{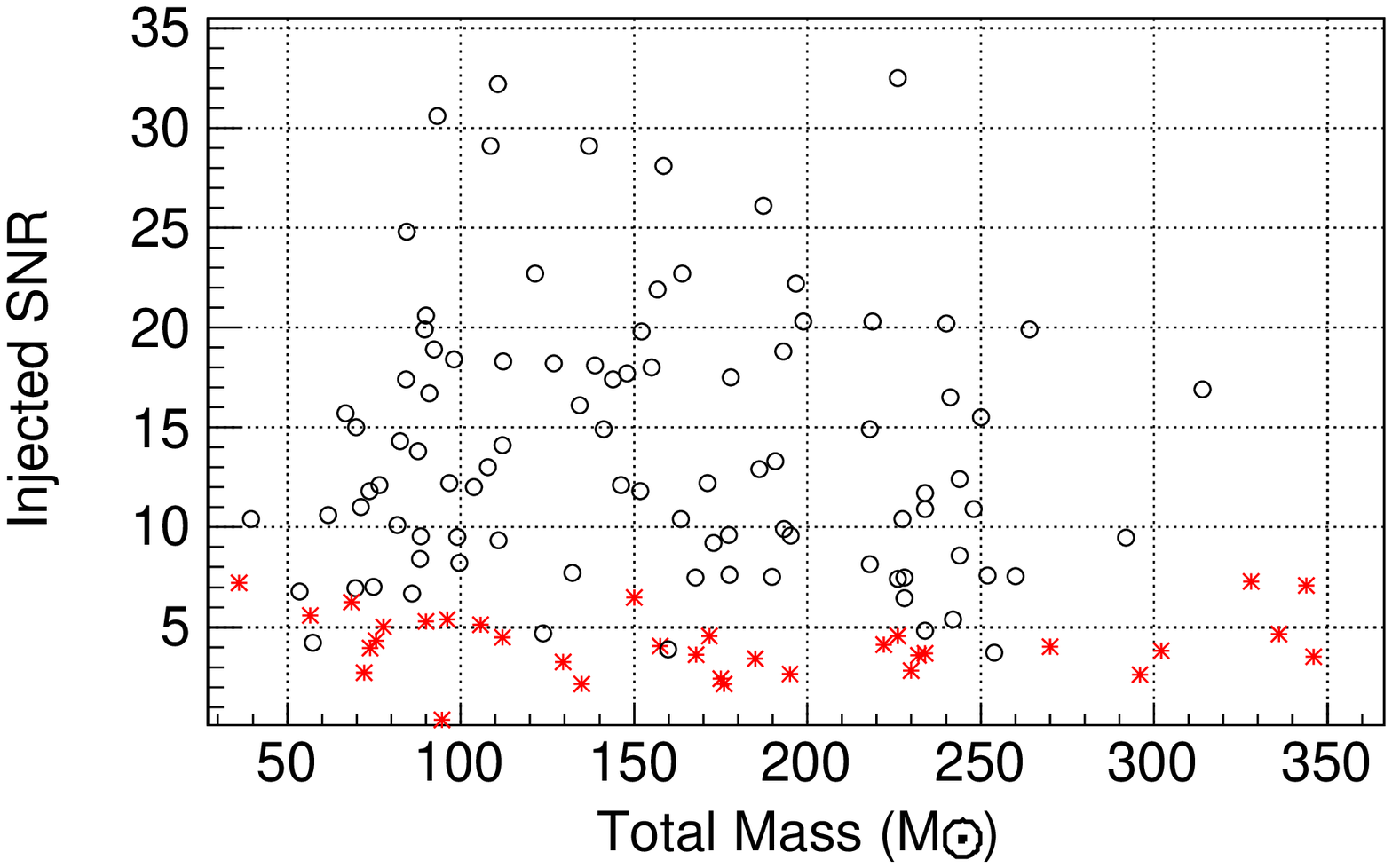}
  \includegraphics[width=0.49\linewidth]{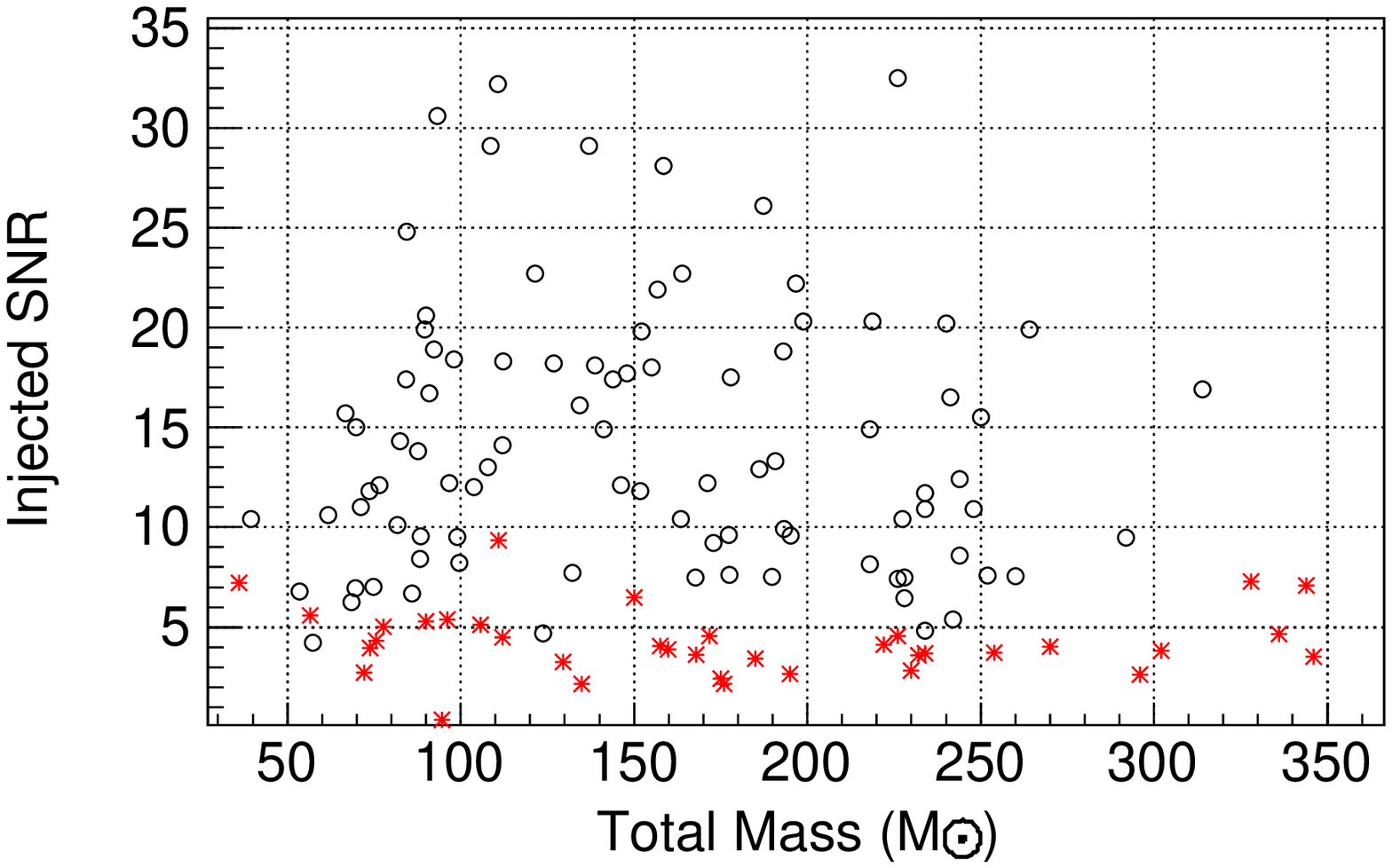} 
  \caption{Injections that were missed (stars) and found (circles) by {\em Q-pipeline} (left) and by {\em lalapps\_ring} (right)
    in H1 as a function of SNR and total mass. For both algorithm, the condition for an injection to be {\it found} is that the measured signal-to-noise ratio is SNR$_\mathrm{measured}\ge 5.5$.}
  \label{fig:missedfound}
\end{figure}

\begin{figure}[htb]
  \centering
  \includegraphics[width=0.49\linewidth]{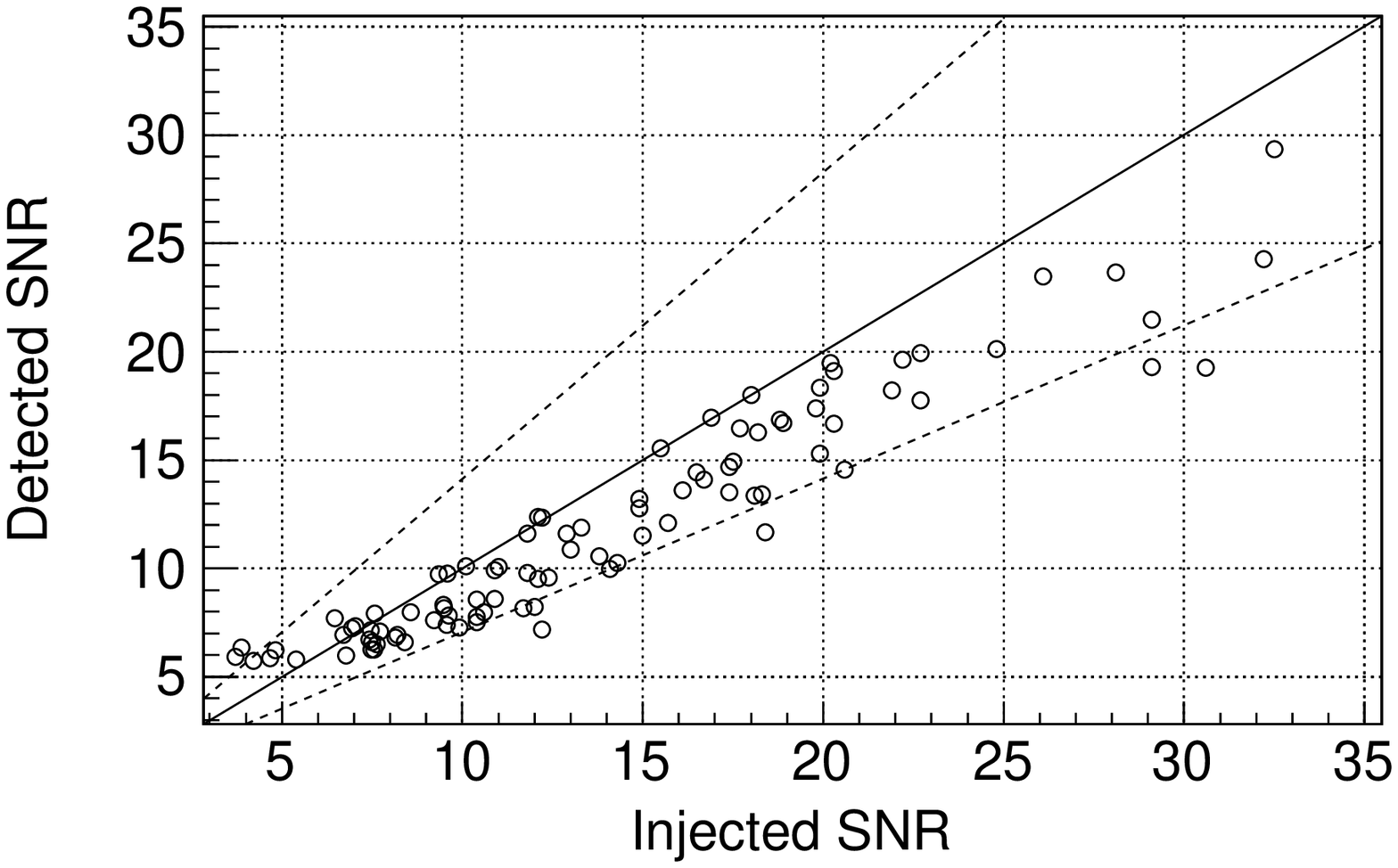}
  \includegraphics[width=0.49\linewidth]{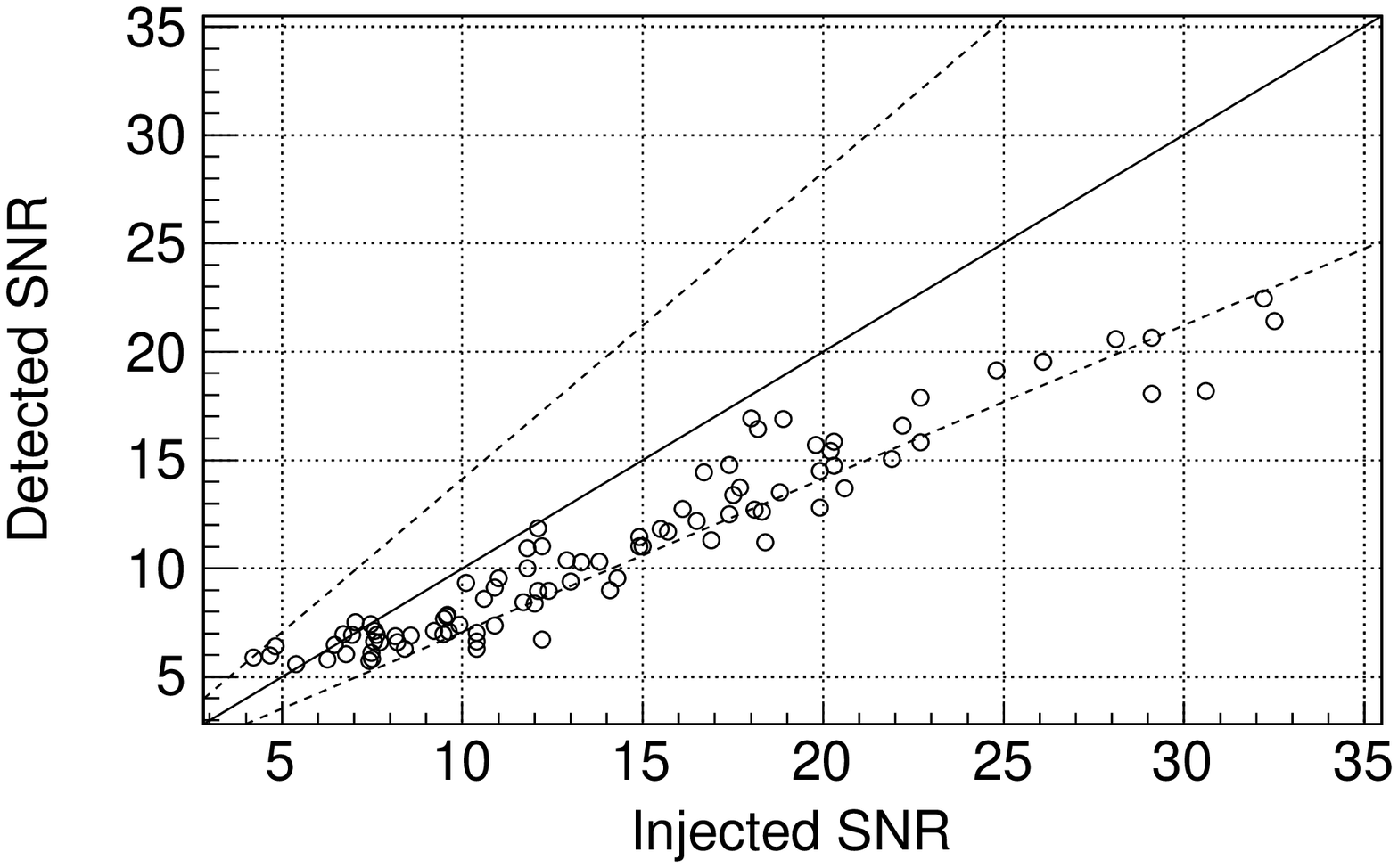}
  \caption{SNR detected by by {\em Q-pipeline} (left) and by {\em lalapps\_ring} (right) as a function of injected SNR for found 
    injections. Dashed lines indicate a deviation from the diagonal by a 
    factor of~$\sqrt{2}$. }
  \label{fig:SNR}
\end{figure}

Fig.~\ref{fig:missedfound} shows plots of all the injections used in the NINJA project as a function of injected SNR and total mass, identifying which were missed and found by {\em Q-pipeline} and {\em lalapps\_ring}; black circles are injections found  with SNR$\ge 5.5$, the red stars are missed. 
Out of 94 signals with injected SNR$\ge 5.5$, 88 (88) were found by {\em Q-pipeline} ({\em lalapps\_ring}) and 89 (87) were found by {\em Q-pipeline} or (and) {\em lalapps\_ring} with measured SNR$\ge 5.5$.  Of the two signals that were not found by both, one was very close to threshold, and the other (stronger) was missed due to the deadtime built in the filter for training purposes.
A comparison of the difference between injected and measured SNRs is in Fig.~\ref{fig:SNR}, where we plot the SNR detected by the {\em Q-pipeline} and {\em lalapps\_ring} as a function of the injected SNR.  
In both cases the detected SNR is smaller than the injected SNR (for SNRs above threshold), which is consistent with the two algorithms only detecting a portion of the coalescence. There are few events at the detection threshold where the detected SNR is larger than injected SNR, due to noise fluctuations near threshold.  
For stronger signals, the discrepancy between measured and injected SNR is roughly within a factor of $\sqrt{2}$, indicated by the dashed lines in Fig.~\ref{fig:SNR}. 
This first comparison needs to be pursued with higher statistics, to understand how the signal-to-noise ratio is affected by the different whitening procedures used by the two algorithms, and by what fraction of the signal is detected. 

\begin{figure}[htb]
  \centering
  \includegraphics[width=4in]{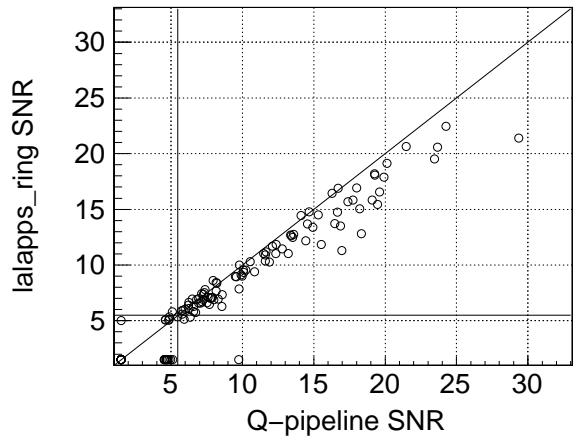}
  \caption{SNR detected by {\em lalapps\_ring} versus SNR detected by {\em Q-pipeline} for injections with injected SNR~$\geq 5.5$.  The solid vertical and horizontal lines mark the threshold SNR of~5.5; an SNR set to~1.5 indicates a missed injection.  Note that Q-pipeline tends to return a slightly higher SNR, consistent with it picking up a larger portion of the waveform than {\em lalapps\_ring}.
    }
  \label{fig:SNRcomp}
\end{figure}

\begin{figure}[htb]
\centering
\includegraphics[width=4in]{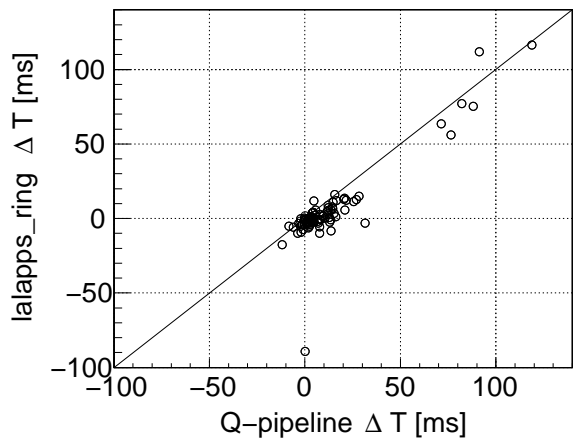}
\caption{Timing accuracy of detected injections by {\em lalapps\_ring} versus {\em Q-pipeline}.  Here $\Delta T = T_\mathrm{meas} - T_\mathrm{peak}$, where $T_\mathrm{meas}$ is the recovered time, which for {\em Q-pipeline} is the central time of the Gaussian envelope, and for {\em lalapps\_ring} is the beginning of the ring-down waveform template.  $T_\mathrm{peak}$ is the time of the waveform maximum, as described in~\cite{Aylott:2009ya}.}
\label{fig:time}
\end{figure}

\begin{figure}[htb]
  \centering
  \includegraphics[width=0.49\linewidth]{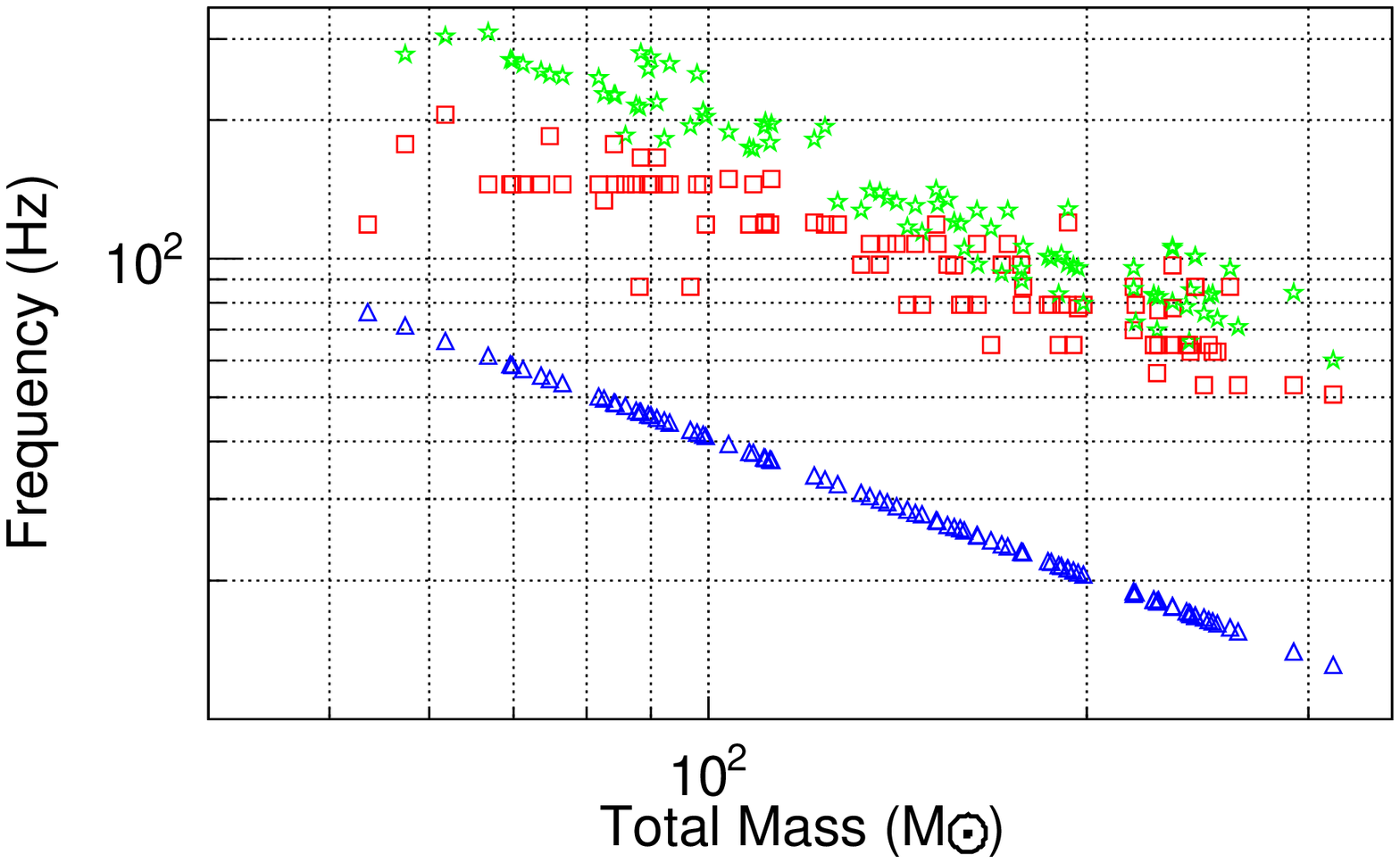}
  \includegraphics[width=0.49\linewidth]{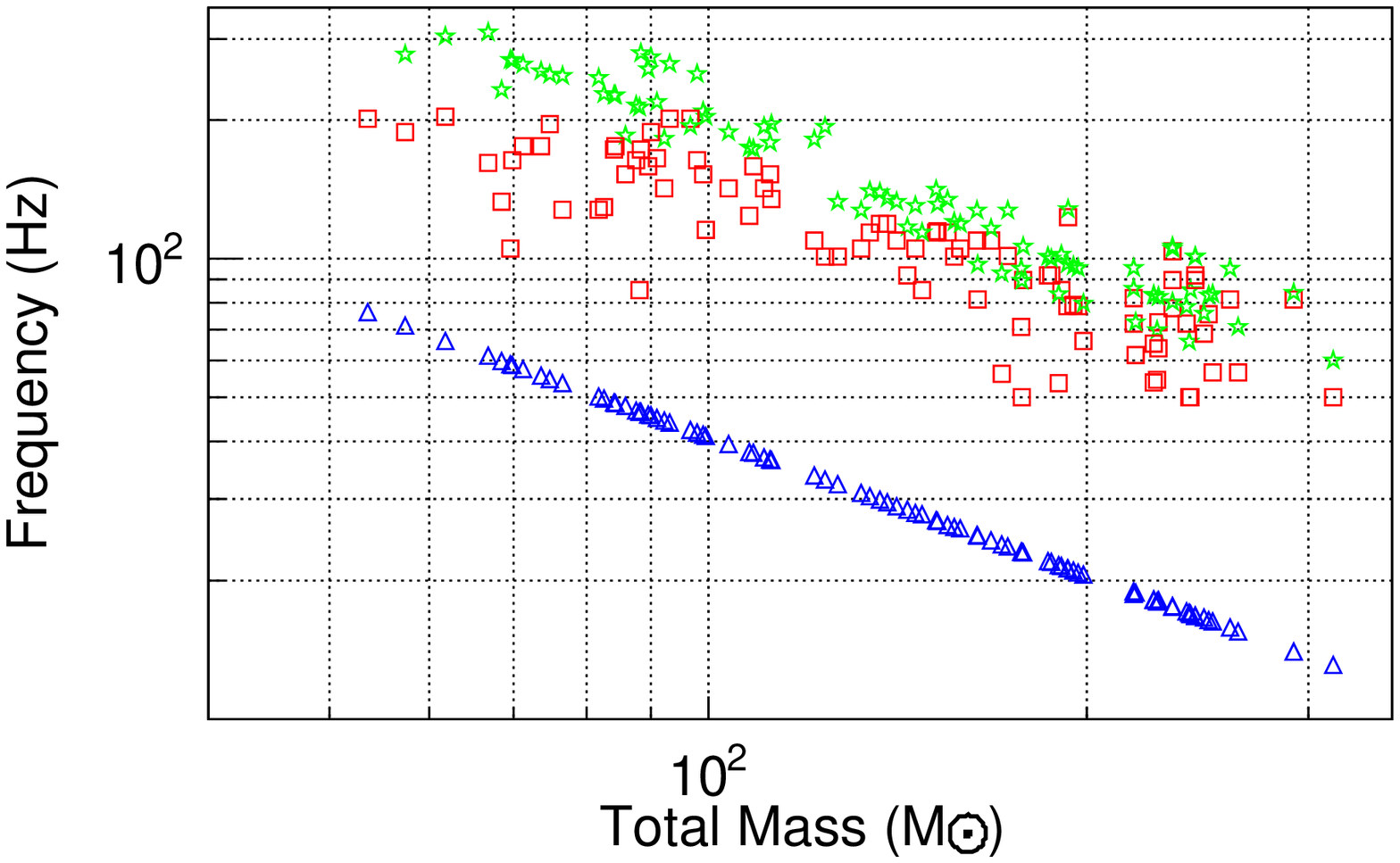}
  \caption{Frequency as a function of mass. The squares are the frequency 
    returned by {\em Q-pipeline} (left) and {\em lalapps\_ring} (right). In both figures, triangles are~$f_\mathrm{ISCO}$; stars are~$f_\mathrm{ring}$, computed from Eq.\ref{e:freq}. The discretization of frequencies in the {\em Q-pipeline} plot is inherited from  the sine-gaussian template bank.}
\label{fig:freqQ}
\end{figure}

\begin{figure}[htb]
  \centering
  \includegraphics[width=4in]{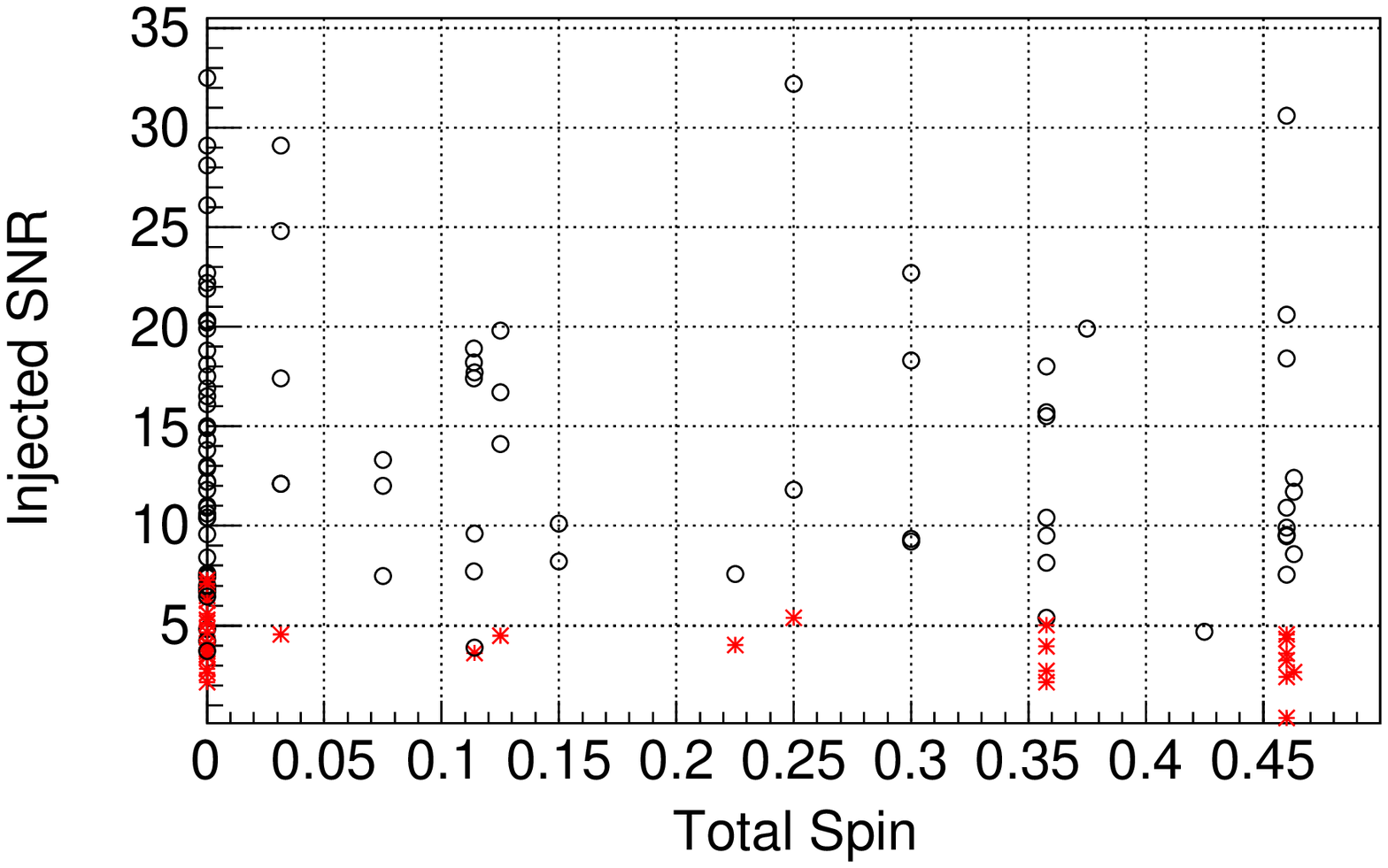}
  \caption{Missed (stars) and found (circles) injections for {\em Q-pipeline} versus modulus of the total spin $a$ measured in dimension-less spin unit 
    %%%$a=\frac{c}{G}\left(\frac{|\vec{S}_1|}{m_1^2}+\frac{|\vec{S}_2|}{m_2^2}\right)$
    $a = \frac{c}{GM^2}\left|\vec{S}_1+\vec{S}_2\right|$, 
    $\vec{S}_i$ being the individual spin vectors. {\em lalapps\_ring} yields the same result.}
\label{fig:spin}
\end{figure}

To compare the detection efficiency of {\em Q-pipeline} with {\em lalapps\_ring}, in Fig.~\ref{fig:SNRcomp} we also plot the SNR recovered by {\em lalapps\_ring} against the one recovered by {\em Q-pipeline}.  
Note that {\em Q-pipeline} finds injections with a slightly larger SNR; however, the detection performance of the two algorithms is relatively consistent.  Here, the same detection threshold SNR~$\ge 5.5$ was used for both searches; in a real search, different thresholds may be needed for the two algorithms, depending on the false alarm rate.

\subsection{Parameter Estimation}

The {\em Q-pipeline} identifies a central time for the detected event, corresponding to the peak time for the sine-Gaussian waveform with largest SNR, while {\em lalapps\_ring} reports as event time the beginning of the ring-down template with best match to the data.
Fig.~\ref{fig:time} shows the deviation of these two time measurements from the peak time of the injected waveform. 
The two algorithms are in agreement, with a single outlier, visible in the plot. This event corresponds to a weak injection, close to threshold; in this case, {\em lalapps\_ring} triggered on a startup transient of the signal, 90 ms before the merger. 

The {\em Q-pipeline} algorithm returns a frequency corresponding to the central frequency of the most significant tile in the time-frequency domain.  
This frequency, as well as the ring-down frequency from {\em lalapps\_ring}, is plotted in Fig.~\ref{fig:freqQ}, along 
with~$f_\mathrm{ISCO}$ and~$f_\mathrm{ring}$, calculated from the injected parameters according to Eq.~\ref{e:freq}, for comparison.  Note that for both algorithms the recovered frequency 
tends to be from the portion of the coalescence waveform in the most sensitive region of the detector (50-200Hz): the inspiral (triangles) for lower masses, and the ring-down (stars) for higher masses. 
This can be well explained as ISCO and ring-down  frequencies are inversely proportional to the total mass of the binary system so, as injection masses increase and frequencies decrease, the portion of the signal falling into the best sensitivity region of the detectors move from the 
inspiral to the ring-down part of the signal, the largest frequency for any given pair of masses and spins.
For the parameters tested in this study, both algorithms detect signals between~$f_\mathrm{ISCO}$ and~$f_\mathrm{ring}$.

Finally in Fig.~\ref{fig:spin} we show the SNR of missed and found injections for {\em Q-pipeline} as a function of the sum of the spins of the constituents the binary system; {\em lalapps\_ring} yields the same result. This plot does not show any dependence on the spin of the black holes, additional simulations with a broader variety of spin magnitudes and orientations will be needed for a more conclusive statement.

\section{Conclusions}

\label{conclusions}
In the context of the Numerical INJection Analysis (NINJA) project, the {\em Q-pipeline} burst search algorithm successfully analyzed numerical relativity BBH coalescence waveforms for a variety of masses and spins in simulated colored Gaussian noise.  The {\em Q-pipeline} single interferometer performance is comparable to matched filtering to ring-down templates, and yields similar arrival time and frequency, and a slightly better SNR. 
In particular, depending on the total mass of the BBH system, both algorithms trigger between $f_{\rm ISCO}$ and $f_{\rm ring}$ over the broad parameter space covered by NINJA.  
We emphasize this is a qualitative statement: the statistics of the NINJA data set is too small for a quantitative, systematic comparison. 
Moreover the absence of the non-Gaussian noise transients typical of real detectors does not allow a realistic estimation of the false alarm rate and threshold settings. 
Systematic studies will be subjects of future NINJA projects.

\section{Acknowledgments}

The authors thank the NINJA collaboration for sharing waveforms and simulated data and for useful discussion in the definition of the problems and interpretation of results. 
This work was supported by grant NSF PHY-0653550 and by the Istituto Nazionale di Fisica Nucleare (INFN). 

\noindent
This paper was assigned number LIGO Document Control Center number P0900068. 

\newpage

\bibliographystyle{iopart-num}
\bibliography{biblio}

\end{document}